\documentstyle[twocolumn,prb,aps,epsf]{revtex}
\begin{document}
\preprint{Version of June 25}
\draft
%********************************************************************

\title{\bf
%\vspace*{-10mm}
%\begin{flushright}
%{\large {\bf PREPRINT:} Applied Physics Report 96-27}
%\end{flushright}
%\vspace*{5mm}
Spatial Structure
in Mode Population
Induced by Coherent Pumping
in a Ballistic Quantum Channel
}
\author{Ola Tageman, L. Y. Gorelik,
R. I. Shekter and M. Jonson
%\\ \small \em Microwave and High Speed Electronics Research Center
%\\ \small \em Ericsson Microwave systems, 431 84 M{\"o}lndal, Sweden
%\\ \small \em and
\\  Department of Applied Physics, Chalmers University of Technology
\\  and G{\"o}teborg University, S-41296 G{\"o}teborg, Sweden }

%\date{June 1996}

%\begin{document}
\maketitle

\begin{abstract}
We predict a spatially varying mode population to appear
in a GaAs/AlGaAs-2DEG  ballistic quantum channel
pumped by a THz-field.
If a resonant coupling
between two modes is suddenly switched on at the
entrance, Rabi oscillations in the mode population will arise.
We propose to use an array of gates in order to simulate
a moving quantum point contact for detecting
the mode population oscillations, since they discriminate
between different modes.
By consecutively activating them we expect to see both 
photovoltaic effects and photoconductive
effects which can easily be distinguished from noise.
\end{abstract}

%********************************************************************
\section{Introduction}
%********************************************************************

Structures of submicron size, created in a two dimensional
electron gas
in a semiconductor heterostructure
using a split gate technique,
have during the last few years been a subject of intensive 
investigations \cite{Beenakker}.
A short electron traveling time $ \tau \approx 10^{-11} $ s,
in combination with
a high sensitivity
of the electron transport to external fields,
qualifies these
systems as possible
ingredients
in high speed electronic devices.
An important feature of such small systems is a pronounced
quantization of electron motion which brightly manifests itself
as a fundamental conductance quantization in a point contact
\cite{Glazman,Wees,Wharam}.
A preservation of phase coherence, being a necessary
condition for such quantization, makes the scattering of electrons
a process of
quantum mechanical nature in which interference
between different scattering events plays a crucial role.
As a result of this interference a number of localization
effects has been
observed in quantum channels, which arise from impurity scattering
of the electrons.
\cite{Glazman}

Resonant interaction with an external electromagnetic field
resulting in electronic transitions between different quantized states
can also be considered as a kind of scattering (electron-photon scattering),
and the question of how to treat the interference between different
scattering events arises
\cite{Gorelik}. In this paper we will show that resonant
interaction of electrons in a quantum ballistic channel with a strong 
electromagnetic field 
results in a characteristic spatial modulation of the charge
distribution.
This phenomenon is a result of a periodically appearing inversion
of the population, and the period is controlled by the intensity
of the electromagnetic field.
This effect is entirely due to the coherent electron dynamics in the
microwave field and can qualitatively be thought of as a set of
intermode transitions similar to Rabi oscillations \cite{Rabi} but
since the
electrons are moving they will appear as transitions in space along
the channel instead of in time.

The existence of a stationary structure of inversely populated domains
is itself an interesting example of how a non-equilibrium electronic
state may crucially influence the DC-conductance of a microchannel.
Since adiabatic quantum point contacts have the capability of selecting
between the different modes available for propagation
\cite{Glazman,Wees,Wharam}
they may serve
as a perfect tool for detecting the distribution of an electron among the
various modes. We will show that a quantum channel
formed by an array of split gates
may simulate a moving point contact, performing an
effective scan of the induced population structure.
As a result of such a scan we expect to see oscillations in the
current through the channel whose period in time is proportional to the
wavelength of the population modulation, which may be continuously tuned by
changing the intensity of the electromagnetic field.
If instead the circuit is left open we expect to see an oscillating
voltage as a result of the scanning, due to photovoltaic effects.
Our estimations show that a relatively small power of the 
electromagnetic field is needed in order to produce the switching
effect discussed above.

%********************************************************************
\section{Spatial Variation in Mode Population}
%********************************************************************

\begin{figure}[htb] \begin{center}\leavevmode
\epsfxsize\hsize
\epsfbox{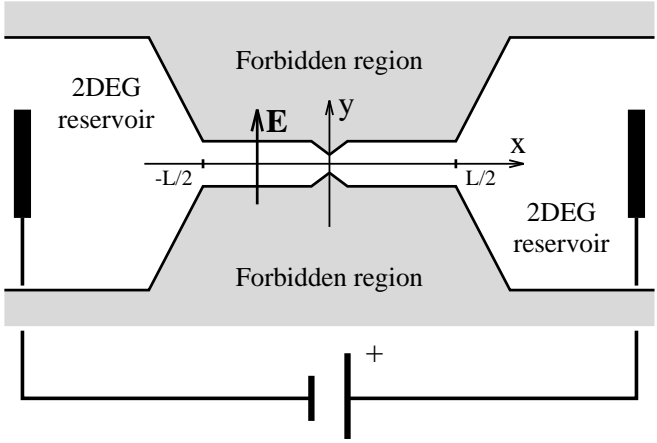}
\caption{
\label{fig:modelsystem}
The two-dimensional model system consists of a narrow channel
connecting two reservoirs.
In the middle of the channel there is a constriction.
The channel region is exposed to a transversely polarized
electromagnetic field.
}
\end{center}
\end{figure}

The basic system under consideration here is presented in figure
\ref{fig:modelsystem}.
Such a structure may be fabricated by placing a
split gate on top of a GaAs hetero structure. A negative voltage on
the gates will confine the electrons to a narrow channel in 
the two-dimensional electron gas (2DEG).
We use a right handed coordinate system with the x-axis along the channel
(pointing to the right) and the z-axis perpendicular to the 2DEG
(out of the plane), as shown in figure \ref{fig:modelsystem}.

We assume the width $d(x)$
to vary smoothly along the channel on a scale of the Fermi wave length,
$\lambda_F \approx 40$ nm, so that we are allowed to use the
adiabatic approximation and hence to separate the transverse from the
longitudinal motion (in absence of external field) \cite{Glazman}.
If the length of the channel, $L$, is small compared to the phase breaking
length, $l_{\phi}$ (caused by inelastic scattering), the charge transport
through the microcontact
can be formulated as a one-particle
quantum-mechanical problem \cite{Landau}.

The interesting dynamics of an electron will arise because of the
coupling to an externally applied high frequency electromagnetic
field, which we assume to be localized in the channel region and
polarized in the y-direction (see figure \ref{fig:modelsystem}).
In this situation the wave function $\Psi(x,y,t)$
of the electron satisfies the
following time dependent Shr{\"o}dinger equation:
\begin{equation}
i\hbar\frac{\partial}{\partial t} \Psi(x,y,t)
=
\hat{H}(t)\Psi(x,y,t)\;,
\label{eq:sequation}
\end{equation}
with the Hamiltonian:
\begin{equation}
\hat{H}(t)=
\frac{1}{2m^*}(\hat{{\bf p}}-\frac{e}{c}{\bf A}(x,t))^2+U(x,y)
\;.
\label{eq:Hamiltonian}
\end{equation}
Here
${\bf p}$ is the momentum,
$m^*$ is the effective mass
of an electron in the 2DEG
and the potential $U$
confines the the electron to the channel
and reservoirs.

For definiteness we will consider $U$ to be a hard wall potential,
given by:
\begin{equation}
U(x,y)=
\left\{
\begin{array}{ll}
0 & \mbox{for $|y|<d(x)/2$} \\
\infty & \mbox{for $|y|>d(x)/2$}
\end{array}
\right.
\;.
\label{eq:hardwall}
\end{equation}

The vector potential $ {\bf A}(x,t)$ in Eq. (\ref{eq:Hamiltonian})
describes an electromagnetic
field of frequency $\omega$ and amplitude $\mbox{\Large $\varepsilon$}$
polarized in the y-direction:
\begin{equation}
{\bf A}(x,t)=\frac{ c}{\omega} \mbox{\Large $\varepsilon$} _{\omega}
f(x) cos(\omega t) {\bf e}_y \;,
\end{equation}
The function $f(x)$ which vanishes in an unspecified way outside
the channel region, $-L/2<x<L/2$, has the effect of focusing the field
to the channel.

Assuming the geometry of the channel to be smooth enough to allow for
adiabatic transport, i.e. assuming $d'(x)\lambda_F/d(x)<<1$, 
it is reasonable to
make the following separation Ansatz for the solution of 
Eq. (\ref{eq:sequation}).
\begin{equation}
\Psi(x,y,t)=\sum_n \Psi_n(x,t)\Phi_n(y,d(x))\;.
\label{eq:separation}
\end{equation}
Here $\Phi_{n}(y,d(x))$ satisfies the equation 
:
\begin{eqnarray}
-\frac{\hbar^2}{2m^*}\frac{\partial^2}{\partial y^2}
\Phi_n(y,d(x))
+
U(y,d(x))\Phi_n(y,d(x))
\nonumber \\
=
E_n(x)\Phi_n(y,d(x))\;,
\end{eqnarray}

For the hard wall potential of Eq. (\ref{eq:hardwall}),
$E_{n}$ and $\Phi_{n}$ are given by:
\begin{eqnarray}
&E_n(x)&=\frac{\hbar^2}{2m^*}\frac{n^2\pi^2}{d^2(x)}
\nonumber \\
&\Phi_n(y,d(x))&=\sqrt{\frac{2}{d(x)}}\sin\left[
\frac{n\pi}{d(x)}\left(y+\frac{d(x)}{2}\right)\right]
\label{eq:modes}
\end{eqnarray}

If we neglect all spatial derivatives of $d(x)$ when inserting the Ansatz
(\ref{eq:separation}) into the Shr{\"o}dinger equation
(\ref{eq:sequation}) we get 
:
\begin{eqnarray}
-\frac{\hbar^2}{2m^*}
\frac{\partial^2}{\partial x^2}\Psi_n(x,t)
+E_n(x)\Psi_n(x,t)+
\nonumber \\
2i\cos(\omega t)\sum_{m}M_{nm}(x)\Psi_m(x,t)
=
i\hbar\frac{\partial}{\partial t}\Psi_n(x,t)\;,
\label{eq:longitudinal}
\end{eqnarray}
Clearly, the transversely polarized electromagnetic field generates
a coupling between different modes. In Eq.
(\ref{eq:longitudinal}) this coupling is described by
the intermode transition elements $M_{nm}$ which for the hard wall
potential of Eq. (\ref{eq:hardwall}) are given by (n+m even gives zero):
\begin{equation}
M_{nm}=\frac{\hbar e\mbox{\Large $\varepsilon$} _{\omega}}{m^*\omega d(x)}
\cdot \frac{2nm}{n^2-m^2}
\equiv V_{\omega}\cdot \frac{3}{2}\frac{nm}{n^2-m^2}\;,
\label{eq:Matrixelement}
\end{equation}

We now consider a situation in which the inter level distance
in the straight part of the channel is much larger than the 
characteristic inter level coupling energy, $V_{\omega}$,
defined in Eq. (\ref{eq:Matrixelement}).
Hence the electromagnetic field can couple two modes, $n$ and $m$
strongly inside the straight part of the channel, only if the
resonance condition $\hbar\omega = |E_{n}^{st}-E_{m}^{st}|$ is 
fulfilled,
where $E_{n}^{st}$ is the transverse energy of
mode $n$ in the straight part of the channel.
Other modes can be taken into account in the
framework of perturbation theory based on the small parameter
$V_{\omega}/|\hbar\omega - E_{n}^{st}-E_{m}^{st}|$

In order to get a dramatic outcome of the mixing we
restrict our attention to the case when
$\hbar\omega = E_{2}^{st}-E_{1}^{st}$ and
$E_{n>2}^{st} > E_{F}$
, where $E_{F}$ is the Fermi energy.
In this case only the first mode enters the channel and
consequently only the two lowest
modes will be involved in the transport of current,
since they are resonantly coupled.
Under these circumstances it seems reasonable that we restrict our 
attention to the $n=1,2$ part of Eq. (\ref{eq:longitudinal}),
and look for a solution in the following form,
\begin{eqnarray}
\Psi^{\sigma}_1(x,t)
=&
\varphi^{\sigma}_1(x)
\Psi^{\sigma}_{1,E}(x)
e^{-\frac{iEt}{\hbar}}
+O (V_{\omega}/\hbar\omega)&
\nonumber \\
\Psi^{\sigma}_2(x,t)
=&
\varphi^{\sigma}_2(x)
\Psi^{\sigma}_{2,E+\hbar\omega}(x)
e^{-\frac{i(E+\hbar\omega)t}{\hbar}}
+O (V_{\omega}/\hbar\omega)&\;,
\label{eq:solution}
\end{eqnarray}
where
\begin{equation}
\Psi^{\sigma}_{n,E}(x)
=
\frac{1}{\sqrt{v_n}}
e^{\frac{i\sigma (S_{1,E}(x)+S_{2,E+\hbar\omega}(x))}{2\hbar}}\;.
\label{eq:WKB}
\end{equation}
In the latter expression
:
\begin{eqnarray}
S_{n,E}(x)
=
\int^{x}_0 dx'\sqrt{2m^*(E-E_n(x'))}
\nonumber \\
v_{n,E}=\frac{1}{m^*}/\frac{\partial S_{n,E}(x)}{\partial x}\;,
\label{eq:action}
\end{eqnarray}
is the classical action and the semi classical velocity
(without external field).
The indices $\sigma =\pm$, defines the two directions of electron
propagation. The plus-sign corresponds to motion from left to 
right and the minus sign indicates motion in the opposite direction.

Substituting the Ansatz (\ref{eq:solution}) into Eq.
(\ref{eq:longitudinal}) one gets an equation for the
mode population amplitudes $\varphi _n^{\sigma} (x)$:
\begin{equation}
i\frac{\partial}{\partial x}\vec{\varphi}^{\sigma}(x)
-P_E(x)\mbox{\boldmath $\sigma$}_z\vec{\varphi}^{\sigma}(x)
+\Lambda_{\omega} \mbox{\boldmath $\sigma$}_y\vec{\varphi}^{\sigma}(x)
=0\;,
\label{eq:popamp}
\end{equation}
where $\mbox{\boldmath $\sigma$}_y$ and
$\mbox{\boldmath $\sigma$}_z$ are Pauli matrices and
\begin{eqnarray}
\vec{\varphi}^{\sigma}(x)
=
\left(
\begin{array}{c}
\varphi^{\sigma}_2(x)
\\
\varphi^{\sigma}_1(x)
\end{array}
\right)
\quad,
\Lambda_{\omega}=\frac{V_{\omega}}{\hbar\sqrt{v_{1,E}(x)
v_{2,E+\hbar\omega}(x)}}
\nonumber \\
P_E(x)=\frac{S'_{1,E}(x)-S'_{2,E+\hbar \omega}(x)}{2 \hbar}
\end{eqnarray}

In general the solution of equation (\ref{eq:popamp})
according to (\ref{eq:separation}) and (\ref{eq:solution})
describes the
coherent spatial mixture of two electron wavefunctions.
One of them corresponds to semiclassical electron
propagation at some energy $E$ in mode 1, and the other
at energy $E+\hbar \omega$ in mode 2.
On the other hand it is clear that the solution which
has physical meaning should correspond to pure transverse
states for the incoming waves.
We will use this for classifying the electron states.
By
$\vec{\varphi}^{\sigma}_{1,E}(x)$
and
$\vec{\varphi}^{\sigma}_{2,E}(x)$
we denote solutions of Eq. \ref{eq:popamp}
which originate from mode 1 at energy $E$ and
mode 2 at energy $E$, respectively.
Explicitly, these solutions correspond to the following
incoming waves $\Psi^{\sigma}_{inc}$ in the
reservoirs:

\begin{eqnarray}
\vec{\varphi}^{\sigma}_{1,E}(x)
\Longleftrightarrow
\Psi^{\sigma}_{inc}
=
\frac{\phi_1(y,d(x))}{\sqrt{v_1(x)}}
\;e^{
i(\sigma S_{1,E}(x)
-\frac{Et}{\hbar})
}
\nonumber \\
\vec{\varphi}^{\sigma}_{2,E}(x)
\Longleftrightarrow
\Psi^{\sigma}_{inc}
=
\frac{\phi_2(y,d(x))}{\sqrt{v_2(x)}}
\;e^{
i(\sigma S_{2,E}(x)
-\frac{Et}{\hbar})
}
\label{eq:classification}
\end{eqnarray}

Note that
$\vec{\varphi}^{\sigma}_{2,E}(x)$
is given by Eq. 
(\ref{eq:popamp})
if $E$ is changed to $E-\hbar \omega$.
The two solutions are related as follows:
\begin{equation}
\vec{\varphi}^{\sigma}_{2,E}(x)
=
\mbox{\boldmath $\sigma$}_y
(\vec{\varphi}^{\sigma}_{1,E-\hbar \omega}(x)
)^*
\end{equation}

In the straight part of the channel $P(x)= P^{st}=$ const,
which gives us the following two linearly independent solutions:
\begin{equation}
\vec{\varphi}^{\sigma}_{\pm}(x)
=
\left(
\begin{array}{c}
\sqrt{1\pm \sin \lambda}
\\
\mp i\sqrt{1\mp \sin \lambda}
\end{array}
\right)
\cdot e^{\pm iKx}\;,
\label{eq:eigensolutions}
\end{equation}
where:
\begin{eqnarray}
K=\sqrt{P^2+\Lambda^2_{\omega}}
\;\;,
%\mbox{\boldmath and}
\;
\sin \lambda \equiv \frac{P}{\sqrt{P^2+\Lambda^2_{\omega}}}
\label{eq:Rabiwavevector}
\end{eqnarray}

The linear combination of the two eigensolutions
(\ref{eq:eigensolutions}) which describes the electron propagation
through the channel is uniquely determined by
the value of the wavefunction at the entrance
of the channel, $\vec{\varphi}^{\sigma}_E(x=\sigma L/2)$.
It is clear that this value depends on how we bring the electron
from the reservoir to the entrance. In the Appendix we show that this
is determined by a relationship between the Rabi wave length, 
$2\pi/K_{st}$ in the straight part
and the size of the transition region $R_{tr}$.
We can distinguish two limiting cases, namely sudden switching,
$K_{st}R_{tr}\ll1$, and adiabatic switching, $K_{st}R_{tr}\gg1$.

In the case of sudden switching we can neglect the influence
of the electromagnetic field on the wavefunction in the entrance
to the channel,
and use the following boundary conditions:
\begin{equation}
\vec{\varphi}^{\sigma}(x=-\sigma\frac{L}{2})
=
\vec{\varphi}^{\sigma}(x=-\sigma\frac{L}{2})\mid _{\varepsilon
_{\omega}=0}
\label{eq:sudden}
\end{equation}

The boundary conditions are more complicated in the case
of adiabatic switching, but we will not pay more attention
to them here since the main results in this paper concern
the case of sudden switching.

There is a natural reason for classifying the switching
as either sudden or adiabatic. Exactly at resonance the
velocity of an electron is the same in the two modes.
Therefore, viewed in the rest frame of the electron, we
see transitions in a two level system which is being brought
into resonance.
If this process of bringing it into resonance is very slow
the system has time to adjust itself
to the external conditions, and we have ``adiabatic switching''.
If it is very fast the system 
has no time to evolve which means that
it will remain in the initial state.

In the following considerations we will assume that the size of the
transition region (in which this process takes place) as well as the
size of the point contact region is much smaller than the Rabi wave
length. This ensures that sudden switching takes place both at the entrance
and at the microconstriction.

One can get,
using the boundary conditions (\ref{eq:sudden}),
the following wave function describing the electronic
distribution inside the straight part of the channel
($|x|<L/2$):
\begin{eqnarray}
&\vec{\varphi}^{\sigma}_{1,E}(x)
=&
t^{tr}_1 (E)
%\nonumber \\
\cdot
\left(
\begin{array}{c}
\cos \lambda \sin(K(x-\sigma L/2))
\\
\cos (K(x-\sigma L/2))-
\\
i\sin \lambda
\sin(K(x-\sigma L/2)) 
\end{array}
\right)
\nonumber \\
&\vec{\varphi}^{\sigma}_{2,E}(x)
=&t^{tr}_2 (E)
\mbox{\boldmath $\sigma$}_y
(\vec{\varphi}^{\sigma}_{1,E+\hbar \omega}(x))^*
\label{eq:wavefunction}
\end{eqnarray}
Here $t^{tr}_{i}$ is the amplitude of electronic transmission
through the transition regions in a field free case.
Due to the
adiabaticity
of the microconstriction
geometry we have: $|t^{tr}_{i}(E)|=\theta (E-E^{st}_{i})$.
We will consider a channel possessing mirror symmetry of
the entrances
so that $t^{tr}_{i}$
is the same in both directions.

Let us now take a look at a situation in
which $E^{st}_{2}>E_{F}>E^{st}_{1}$.
In this case the first mode only is responsible for electron
transport whithout the electromagnetic field.
According to Eq. (\ref{eq:wavefunction})
in this situation,
the difference in population
numbers
$\Delta  \rho^{\sigma}_{E} \equiv
(\vec{\varphi}^{\sigma}_{1,E}
,
\mbox{\boldmath $\sigma$}_z
\vec{\varphi}^{\sigma}_{1,E}
)$
is given by:

\begin{equation}
\Delta  \rho^{\sigma}_{E}=
\cos^2 \lambda
\cos (2K(E) (x+\sigma L/2))
\label{eq:difference}
\end{equation}

The periodic variation of mode population mentioned in the
introduction is apparent from expression (\ref{eq:difference}).
Qualitatively the phenomenon can be understood in the following way.
In the rest system of the electron the sudden switching-on of the
resonance field gives rise to Rabi oscillations in the two level
system with a frequency $\Omega_{R}=V_{\omega}/\hbar$.
Since the electron is moving with velocity ${\em v}$
the oscillations will appear in space with a wave vector
$K=\Omega_{R}/{\em v}$.

We may safely predict that the existence of such a spatial
modulation in space will manifest itself in a lot of physical phenomena.
For example, since the transverse charge density is different in 
the different modes we expect to see induced quadrupole moments as
a consequence of the oscillating mode population.
In order to see how this couples back to the transport properties
we must of course take electron-electron correlations into account.
However this is beyond the scope of this work and will be treated
elsewhere.
Below we will demonstrate the manifestation of the mode population
in electron transport through the microconstriction
in the middle of the channel.

%*********************************************************************
\section{Charge transport}
%*********************************************************************

Following the by now standard approach of Landauer \cite{Landauer}
we formulate the electronic transport problem in terms of a
one-particle transmission problem. According to this approach the
general formula for calculating the current through the
microcontact, at zero temperature
can be written as:
\begin{eqnarray}
I=\frac{2e}{h}[ \int_0^{E+\frac{eV}{2}}dE\sum_n P_n^+(E)
\nonumber \\
-\int_0^{E-\frac{eV}{2}}dE\sum_n P_n^-(E)]\;,
\label{eq:current}
\end{eqnarray}
where:
\begin{equation}
P_n^{\pm}(E)=\sum_{ml}|T_{nm}^\sigma (E,E+\hbar \omega l)|^2\;.
\label{probamps}
\end{equation}
Here, $T_{nm}^\sigma (E,E+\hbar \omega l)$
is the probability amplitude for an intermode electronic transition
resulting in absorbtion or emission of $l$ quanta of the 
electromagnetic field.
In expression (\ref{eq:current}), $V$ is the driving voltage
which is responsible for the difference in chemical potential between
the two reservoirs.

We will consider a situation in which only the lowest two modes,
$n=1$ and $n=2$, which are assumed to be strongly coupled, contribute
to the electronic transport. Therefore we
focus on the following four kinds of electronic transitions:
\begin{eqnarray}
\begin{array}{lr}
\begin{array}{l}
(1,E) \rightarrow (1,E)
\\
(1,E) \rightarrow (2,E+\hbar \omega)
\end{array}
&
\begin{array}{l}
(2,E) \rightarrow (2,E)
\\
(2,E) \rightarrow (1,E-\hbar \omega)
\end{array}
\end{array}
\nonumber
\end{eqnarray}

If there are no back scattering processes in the channel
we find, using Eq. (\ref{eq:wavefunction}), the
transition amplitudes and probabilities for these
processes to be:
\begin{eqnarray}
|T_{nm}^\sigma (E)|^2=
|\varphi_{n,E;m} (x=\sigma L/2)|^2 \cdot |t_n^{tr}(E)|^2
\nonumber \\
P_n=\sum_m|T_{nm}(E)|^2=|T_n^{tr}(E)|^2=\Theta(E-E_n^{st})\;.
\label{eq:noback}
\end{eqnarray}
In these equations we have omitted the second energy argument
of $T_{nm}(E,E+\hbar \omega l)$ since it is implicitly given
by the mode indices $n$ and $m$ together with the initial
energy $E$. We have also used the fact that the solution of
Eq. (\ref{eq:popamp}) satisfies the condition:
$|\vec{\varphi}|^2=$ const, which is a consequence of
charge conservation.

It is clear from an inspection of Eq.s (\ref{eq:current})
and (\ref{eq:noback}) that the total current is given simply
by the probabilities $|t_n^{tr}(E)|^2$ to reach the entrance
of the channel for an electron in mode n.
In the ``sudden switching'' regime, these probabilities
are independent of the electromagnetic field
and therefore in this case, a transversely
polarized field does not at all affect the charge transport 
in absence of backscattering inside the channel.

\begin{figure}[htb] \begin{center}\leavevmode
\epsfxsize\hsize
\epsfbox{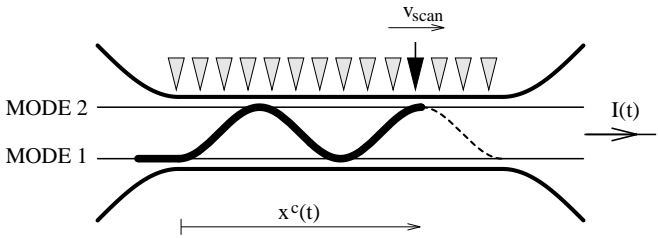}
\caption{
\label{fig:scanning}
A simplified description of the mechanism for scanning
the mode population structure is shown. The electromagnetic
field produces population oscillations between two modes
along the channel. By consecutively activating one gate at
a time we effectively move a filter that stops  mode 2,
and thereby generates an oscillating
current in the channel.
}
\end{center}
\end{figure}

The situation however changes drastically if we put a micro
constriction in the channel at a point $x=x^c$ which enables
back scattering processes to appear. Such a microconstriction
is created when we apply a negative voltage on one of the
gates along the channel as indicated
schematically in figure \ref{fig:scanning}.

The reason for having more than one gate
is that it provides a possibility to
control the $x$-position of the microconstriction.
We will assume that the extension of the microconstriction is
much smaller than the Rabi wave length and also that the intermode
distance $E_2^c-E_1^c$ is much larger than the electromagnetic
energy quantum $\hbar \omega$, where $E_n^c$ is the transverse
energy in the most narrow part of the constriction.
These assumptions
justify a treatment of the transport properties
of the microconstriction which neglects the influence of the
electromagnetic field in this particular part of the channel.
In addition they justify a sudden switching approach
in the transition region between the straight
and the constricted part of the channel.

We therefore consider the total transmission as a result of
the following (sequentially appearing)
three transmission processes. 
Firstly, there is an inelastic 
transmission process from the entrance
of the channel to the microconstriction, secondly, there is
an elastic transmission process through the microconstriction,
and finally there is an inelastic transmission process from the
microconstriction
to the exit of the channel.
It should be noted that the last of these processes
cannot influence the current transport since it does not affect
the total transmission probability, and therefore
we do not have to bother about it at all.
We end up with the following expressions for the total probability
for transmission in the two modes and the two directions:
\begin{eqnarray}
&P_1^{+(-)}=|T_1^{tr}(E)|^2\times&
\nonumber \\
&(|T_{11}^{L(R)}(E)|^2 p_1(E)&+
|T_{12}^{L(R)}(E)|^2 p_2(E+\hbar \omega))
\nonumber \\
&P_2^{+(-)}=|T_2^{tr}(E)|^2\times&
\nonumber \\
&(|T_{22}^{L(R)}(E)|^2 p_1(E)&+
|T_{21}^{L(R)}(E)|^2 p_1(E-\hbar \omega))
\;,
\label{eq:transprocesses}
\end{eqnarray}
Here $T_{nm}^{L(R)}$ is the scattering amplitudes for the
left (right) part of the channel, and $p_n(E)=
\Theta (E-E_n^c)$ is the transmission
probability of the
elastic transmission process trough the microconstriction.
The Eqs. \ref{eq:current}, \ref{eq:transprocesses},
\ref{eq:noback} and
\ref{eq:wavefunction}
completely determine the current through the channel
containing the microconstriction exposed by a resonance
electromagnetic field.

%*********************************************************
\section{Photo conductance}
%*********************************************************

Let us for a start consider a symmetric geometry
which means that the microconstriction is located in the
middle of the channel. In this case $|T^R|=|T^L|$
(and $P^+=P^-$) and there
is no current without a driving voltage applied.

In order to highlight a particularly interesting phenomenon
we will assume that the following 
inequalities are satisfied: $E_2^{st}>E_F>E_1^c,
E_2^{st}<E_F+\hbar \omega<E_2^c$.
These inequalities are fulfilled by a proper choice of gate
voltage (on both the electrodes forming the channel and on
the electrodes forming the microconstriction).
Two things are achieved
by this choice
Firstly, at the Fermi energy the first mode only, is propagating
in the channel, and it is even propagating in the microconstriction.
Secondly, at an energy $E_F+\hbar \omega$ also the second
mode is propagating in the channel, but not in the
microconstriction. This is the key to the mode selection
mechanism.

\begin{figure}[htb] \begin{center}\leavevmode
\epsfxsize\hsize
\epsfbox{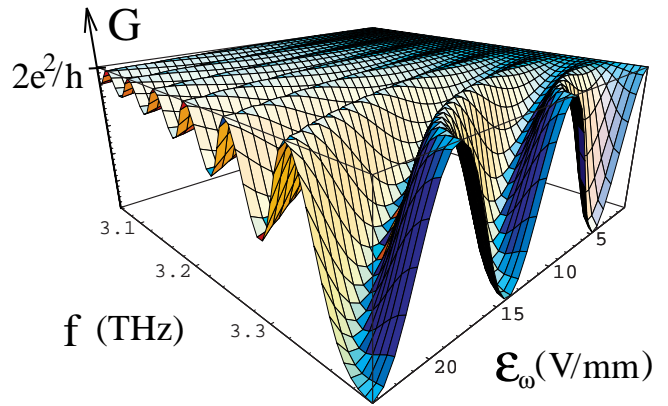}
\caption{
\label{fig:freqfield}
A plot of the conductance versus
the frequency and the strength of the external field.
Note the resonant
character and
the periodic dependence on the fieldstrength.
The resonance frequency ($\approx 3.4$ THz) constitutes
a symmetry line.
}
\end{center}
\end{figure}

Combining the above inequalities with Eq.
(\ref{eq:transprocesses}) and Eqs.
(\ref{eq:wavefunction})-(\ref{eq:noback}) 
we find the
following expression for the conductance
\begin{eqnarray}
G=\frac{2e^2}{h}
(1-\Delta \rho_{E_F}(x=0))
\nonumber \\
=\frac{2e^2}{h}(cos^2(K(E_F)L/2))
\nonumber \\
+
\frac{P^2}{P^2+\Lambda^2}
sin^2(K(E_F)L/2))
\;,
\label{eq:transmission}
\end{eqnarray}
{}From this expression we see that in the case of perfect
resonance ($P=0$) the current is completely blocked
if $K(E_F)L=V_0L/\hbar v_1(E_F)=\pi(2n+1)$.
The conductance as a function of both frequency, $f$ and
field strength, $\mbox{\Large $\varepsilon$}_\omega$
of the electromagnetic field is plotted in figure
\ref{fig:freqfield}, for the case: $d=35 $ nm, $L=5$ $\mu$m and $E_F=14$ meV.

%*********************************************************
\section{Photovoltaic effect}
%*********************************************************

Let us now study the electron transport when we form
a microconstriction not in the middle of the channel
but at a point $x=x^c\neq0$. It is known that if the
mirror symmetry is broken a photo current
will flow even without of a driving voltage
\cite{Nazarov,Vyurkov}.

In the constant current regime
the photocurrent gives rise to a compensating voltage
difference between the reservoirs
\cite{Nazarov,Vyurkov}.
Thus we have a photovoltaic effect
in the system. According to Eqs. \ref{eq:current},
\ref{eq:noback} and \ref{eq:transprocesses} in strong
resonance, in the zero current
regime, this photo voltage is
found from the following equation:
\begin{equation}
eV_{ph}=
\sum_\sigma \sigma
\int_{E_1^c}^{E_F+\sigma eV_{ph}/2}
\Delta \rho ^{\sigma}_E (x^c)dE
\label{eq:zeroregime}
\end{equation}

\begin{figure}[htb] \begin{center}\leavevmode
\epsfxsize\hsize
\epsfbox{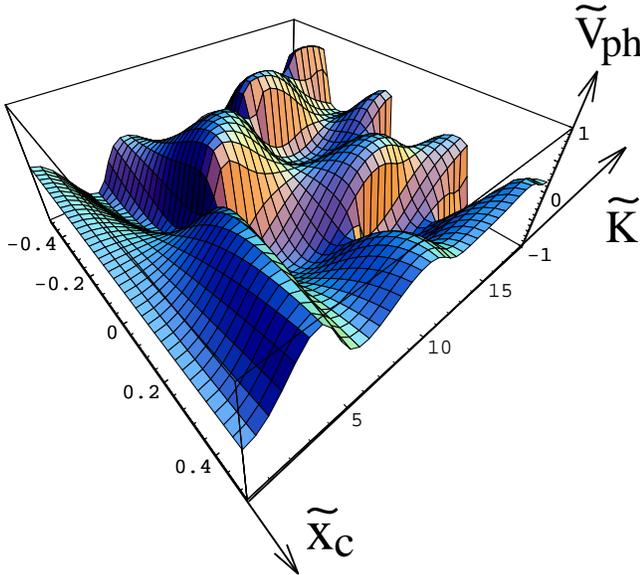}
\caption{
\label{fig:compphoto}
A plot of the normalized photovoltage
versus the normalized Rabi wave vector and
the normalized location of the microconstriction
which is generated between the two reservoirs
if the circuit is left open.
}
\end{center}
\end{figure}

According to Eq. (\ref{eq:difference})
this gives a transcendental equation for $V_{ph}$.
A plot of the normalized photovoltage
$\tilde{V}_{ph} \equiv eV_{ph}/(E_F-E_1^{st})$  versus the normalized
Rabi wave vector $\tilde{K} \equiv K(E_F)L$ and
the normalized location of the microconstriction
$\tilde{x_c} \equiv x_c/L$ for
the typical case $E_F-E_1^{st}=2(E_F-E_1^c)$ is shown in figure
\ref{fig:compphoto}. 

{}From Eq.
\ref{eq:zeroregime}
it is clear that the photovoltage
depends on the position of the microconstriction,
the amplitude of the microwave field (via the Rabi
wave vector $K$) and the width of the microconstriction
(via $E_1^c$). By manipulating these two
parameters one can reach a state in which $K(E_F)L=\pi n$
and $K(E_1^c)L=\pi(m+\frac{1}{2})$. In such a state,
at $m\gg 1$, the
photovoltage is given by:
\begin{eqnarray}
V_{ph}=\frac{2(E_F-E_1^{st})}{eK(E_F)L}
\sin(2K(E_F)x^c)
=
\nonumber \\
\frac{2(E_F-E_1^{st})}{eK(E_F)L}
\sqrt{1-\Delta \rho_{E_F}(x^c)}
sign(\Delta \rho_{E_F}(x^c)
\;,
\label{eq:statevoltage}
\end{eqnarray}
where we have omitted the superscript of $\Delta \rho$
because when $KL=\pi n$ we have $\Delta \rho^+=
\Delta \rho^-$.

Eq. (\ref{eq:statevoltage}) gives the photo voltage
as a function of the population difference $\Delta \rho_{E_F}$
at the point $x=x^c$ and may form a basis for
detecting the intermode population structure in 
the channel. In addition it seems attractive in 
the application perspective. 
For example, we can change the polarity of the photovoltage
by changing the position of the
microconstriction.

As an extension of this example we may scan the superstructure
by moving the gate position with a constant velocity, $x_c=s't'$.
This will result in a photo voltage oscillation
with a frequency proportional to the scanning velocity
and the amplitude of the high frequency field.

%*********************************************************
\section{Discussion}
%*********************************************************

In the introduction we mentioned briefly a scanning procedure
which could be used for investigating the population structure in
the channel. We will here develop these ideas a bit further.
The basic reason for bothering about the scanning
procedure at all is that it will provide a periodic signal
which more easily can be distinguished from noise.
In addition it will present a more complete picture
of the periodic structure in the channel than if fixed
gates are used.

The closest we can come to a continuously moving QPC is to 
have a number of gates uniformly distributed along the
channel and then to activate these in sequence, one at a time,
in a repetitive manner.
During such a scan,
the time development of the current through the channel
will reflect the periodic structure in the channel.
If the gates at one instance happens to be activated in 
a domain of population inversion the current will be more
or less blocked
and we expect to see a minimum in the registered signal.
Figure \ref{fig:scanning} describes this situation schematically.

After some time we have instead activated the gates at a location
of a normal population which results in a maximum in the
registered signal. We may comment on two difficulties concerning
the scanning. Firstly, if the maximum length of the channel is
not an integer number of periods the current signal will be
distorted by a discontinuity. Secondly, in order to avoid aliasing 
in the discretisation process, 
we must have more than two gates for each period of
the population structure in the channel.
The discontinuity will generate a spectrum which is unaffected
by a change in the intensity of the electromagnetic field and
can therefore be distinguished from the frequency that
corresponds to the population structure since this is inversely
proportional to the intensity of the field.

A potentially serious restriction is associated with the single electron
picture used in our discussion. In order to be able to neglect
many-body effects the  parameter $\alpha \equiv e^{2}/\varepsilon\hbar v_{F}$,
which characterizes the relative strength of electron correlation effects, 
has to be small. ($\alpha$ measures
 essentially the ratio between the Coulomb energy
and the kinetic energy of an electron). For $\alpha$ to be small
we should consider a microconstriction with a
reasonably large electron sheet density $N_{s} \geq 10^{11}$ cm$^{-2}$.  This
is quite possible in gate controlled structures (see for example
ref. \onlinecite{lindelof}). 

All effects discussed in this paper are manifestations of
phase coherence even though the mesososcopic systems considered have been 
driven far from equilibrium. Therefore we have to require that phase
breaking processes are ineffective during the time $t_{tr} \simeq L/v_{x}$ it
takes for an electron to pass through our mesoscopic system. Even in
 the absence of
many-body effects ($\alpha \ll 1$) 
interactions give rise to a finite 
relaxation time $\tau_{ee}$ (but does not 
renormalize the entire electron energy spectrum). 
The value of $\tau_{ee}$ can be estimated from Fermi's golden 
rule
and
the criterion $t_{tr} < \tau_{ee}$ translates into the
requirement that 
\begin{equation}
L< d \sqrt{n} \alpha^{-2} ,
\end{equation}
where $n$ is the number of propagating modes.
Taking the assumption $\alpha \ll 1$ into account, 
this inequelity can be easily
fulfilled even for a single propagating mode ($n \simeq 1$) without violating
the condition $L \gg \lambda_{F}$ that ensures that the transport is
adiabatic. 
The geometric resonance discussed in this paper develops if $L>
v_{F}/\Omega_{R} \simeq \hbar  v_{F}/V_{\omega}$ so in order to fulfill both
criteria one needs $V_{\omega}>\Delta E \alpha^{2}$.
A
simple estimation shows that ${\cal E}_{\omega} \simeq$ 100~V/cm is a
sufficient field strength. Such fields are readily available in
real experiments \cite{kollberg}.

In conclusion our estimations show that for semiconductor structures 
with weakly
correlated electrons the effects discussed in this article should be
 experimentally observable. For structures with low electronic densities 
($\alpha \geq 1$) a separate analysis is needed.

Electron-phonon relaxation seems to be less essential for phase breaking than
relaxation due to electron-electron interactions. The corresponding
relaxation frequency 
\begin{equation}
\frac{1}{\tau_{e-ph}} \simeq \omega_{D} (\frac{1}{\lambda_{F} q_{D}})^{3}
\end{equation}
appears to be smaller then $1/\tau_{ee}$ since 
$1/\lambda_{F} q_{D} \simeq 10^{-2}$ for realistic structures.
 
To the extent that it leads to mode mixing, impurity scattering is also a
relevant mechanism that may interfer with the possibility to observe the
photoconductance we have predicted. Impurity potentials are typically weakly
screened and can therefore only give rise to small momentum transfers. If only
a few modes are populated in a mesoscopic structure intermode scattering is
associated with large momentum transfers and are therefore exponentially
suppressed \cite{Yuri}. If many modes are populated, on the other hand,
impurity scattering becomes more effective since it is associated with small
scattering angles.

There are relatively few experimental papers that study
the photoconductance of quantum point contacts; certainly they are few compared
to what has been reported for classical point contacts. Early work clearly
demonstrated that irradiation could influence the conductance of a
quantum point contact via heating \cite{Wyss1,Wyss2}. Only recently have
experiments been performed where convincing evidence for a photoconductance not
caused by heating has been reported \cite{Pepper1,Pepper2}.

%******************************************************************
\section{Acknowledgement}
%******************************************************************
We acknowledge financial support from the Swedish Research Council
for Engineering Sciences (TFR). Ola Tageman is supported also by
Ericsson Microwave Systems.

%*********************************************************************
\section{Appendix}
%*********************************************************************

When calculating the transmission amplitudes corresponding to the
transition region between the reservoir and the entrance to
the channel one can develop a perturbation procedure for the
kinetic part of equation \ref{eq:popamp}
taking into account the smallness of 
the parameter $d'(x)\lambda_F/d(x)\ll1$ (which reflects adiabaticity).
In this procedure we look for a solution to Eq. \ref{eq:popamp}
of the form:
\begin{eqnarray}
\vec{\varphi}^{\sigma}(x)
=&
\vec{\varphi}^{\sigma}_{\pm}(x,P(x))&
+\varepsilon ^{\pm}
\vec{\varphi}^{\sigma}_{\mp}(x,P(x))
e^{\pm i2 \int_0^x K(P(x))dx}
\nonumber \\
&+O(\varepsilon ^2)&
\label{eq:perturbed}
\end{eqnarray}

Here $\vec{\varphi}^{\sigma}(x,P(x))$ is given by Eq.
(\ref{eq:eigensolutions}) if $P$ and $\Lambda$ are replaced
by $P(x)$ and
$\Lambda (x)$ and $Kx$ is replaced by $\int_0^x K(P(x))dx$.
The expansion parameter $\varepsilon^{\pm}$ is supposed
to be small. Substituting Eq. \ref{eq:perturbed}
into Eq. \ref{eq:popamp} we get:
\begin{equation}
\mid \varepsilon ^{\pm} \mid
= \frac{\Lambda'(x)}{2K}
\label{eq:parameter}
\end{equation}

As long as the perturbation procedure is valid, which it is
if $ \mid \varepsilon ^{\pm} \mid \ll1$, the electron will
be switched adiabatically. From Eq. (\ref{eq:parameter}) and
Eq. (\ref{eq:popamp}) and assuming that 
$|\Lambda'(x)|<|P'(x)|$ we get the following condition
for adiabatic switching:
\begin{equation}
|P'(x)|\ll2K^2 \;.
\label{eq:condition}
\end{equation}

Let us now look at the problem in an other aspect. Eq.
\ref{eq:popamp} can be rewritten as:
\begin{eqnarray}
\vec{\eta}\;'(x)&=&
\Lambda \hat{R}\eta(x)
\nonumber \\
\vec{\eta} (x)&=&e^{-i\mbox{\boldmath $\sigma$}_z S_-(x)}\vec{\varphi}(x)
\nonumber \\
\hat{R}(x)&=&
\mbox{\boldmath $\sigma$}_x \cos (2 S_-(x)) +
\mbox{\boldmath $\sigma$}_y \sin (2 S_-(x))
\nonumber \\
S_-'(x)&=&P(x)\;.
\label{eq:etaprime}
\end{eqnarray}

Note that according to Eq. \ref{eq:solution}, $\eta_{1,2}$
are the same as the amplitudes of the wave functions
in the semi-classical case without
a field.
Since $|\eta_2(-\infty)|=0$ and $|\eta_1(-\infty)|=1$
a resonable requirement for preserving the
population troughout the transition region is: $ |\eta_2(0)|\ll 1$.\
Here  we put the zero of
the x-axis at the point were the straight part of the channel begins.

We can simulate the gate geometry of the transition region as
$d(x)=d-\alpha x \Theta (-x)$. 
In this case $|P'(x)|< |P'(0)| \equiv 2R^{-2}_{tr}
\approx \frac{\alpha}{d^2}$.
Near the entrance we have:
$S_-(x)=|P'(0)x^2/2|=R^{-2}_{tr} x^2$.
Assuming $ |\eta_2(x)|\ll 1$ we may take $ \eta_1(x)\approx 1$
and therefore find from Eq. (\ref{eq:etaprime}):
\begin{equation}
|\eta_2(0)|
\approx | \Lambda_{st} \int_{-\infty}^0 e^{i2x^2/R_{tr}^2} dx|
< \Lambda_{st} R_{tr}
< K_{st} R_{tr}
\label{eq:etachange}
\end{equation}
Consequently if $K_{st} R_{tr} \ll 1$ the initial conditions
appropriate for sudden switching are fulfilled.

\end{document}